\documentclass[showpacs,amsmath,amssymb, prb,twocolumn]{revtex4}
\usepackage{graphicx}
\usepackage{dcolumn}
\usepackage{float}
\usepackage{bm}
\begin{document}

\title{Bardeen-Cooper-Schrieffer interaction as an infinite-range Penson-Kolb pairing mechanism}

\author{Francesco Romeo$^{1,2, \ast}$ and Alfonso Maiellaro$^{1,2}$}
\affiliation{$^{1}$Dipartimento di Fisica ''E. R. Caianiello'', Universit\`a degli Studi di Salerno,
Via Giovanni Paolo II, I-84084 Fisciano (Sa), Italy\\
$^{2}$INFN, Sezione di Napoli, Gruppo collegato di Salerno,
Via Giovanni Paolo II, I-84084 Fisciano (Sa), Italy\\
$^{\ast}$email: fromeo@sa.infn.it (corresponding author)\\
\texttt{http://orcid.org/0000-0001-6322-7374}}

\begin{abstract}
We demonstrate that the well-known $(k\uparrow, -k\downarrow)$ Bardeen-Cooper-Schrieffer interaction, when considered in real space, is equivalent to an infinite-range Penson-Kolb pairing mechanism coexisting with an attractive Hubbard term. Driven by this discovery and aiming at exploring the conduction properties, we investigate the dynamics of fermionic particles confined in a ring-shaped lattice. We assume that fermions are simultaneously influenced by the pairing interaction and by an Aharonov-Bohm electromagnetic phase, which is incorporated into the model in a highly non-trivial manner. Remarkably, the aforementioned model shows Richardson integrability for both integer and half-integer values of the applied magnetic flux $\Phi/\Phi_0$, thus permitting the exact solution of a genuine many-body problem. We discuss the ground state properties of both two-particle and many-particle systems, drawing comparisons with results from the attractive Hubbard model. Our approach combines exact diagonalization, density matrix renormalization group techniques, and numerical solution of the Richardson equations. This comprehensive analysis allows us to study various key metrics, including the system's conductivity as a function of the interaction strength. In this way, the BCS-BEC transition is investigated in a continuous manner, thus permitting to shed light on fundamental aspects of superconducting pairing. Our findings can be experimentally tested in a condensed matter context or, with greater level of control, using \textit{atomtronics} platforms.
\end{abstract}
\maketitle
\section{Introduction}
Understanding the emergent properties of interacting fermionic particles remains a fundamental challenge in theoretical physics, with profound implications across a wide spectrum of phenomena extending from the quark structure of baryons\cite{schuckBook} to the superfluid phases in neutron stars\cite{neutronstars}. This endeavor is equally significant in condensed matter physics, where Mott insulators\cite{mott} and superconductors\cite{degennesBook} serve as paradigmatic examples of systems where many-body interactions are responsible for the emergence of entirely new states of matter.\\
Interestingly, while Mott physics is understood in terms of a real space formulation based on repulsive Hubbard interaction among fermions, the microscopic theory of superconductivity due to Bardeen, Cooper and Schrieffer, the so-called BCS theory\cite{cooper56,bcs1,bcs2}, is instead formulated in momentum representation, thus requiring the translational invariance of the system.\\
Shortly after the advent of BCS theory, Russian physicists swiftly grasped its key concepts through the powerful method of the Bogolyubov canonical transformation\cite{bogolyubov}, or by reformulating the theory using the quantum field theory methods pioneered by Gor'kov\cite{gorkov1,gorkov2}. In the successful effort of demonstrating that the Ginzburg and Landau's phenomenological theory\cite{GL} of superconductivity is compatible with the BCS formulation, L. P. Gor'kov introduced a contact interaction in real space (see Ref. [\onlinecite{gorkov1}]) playing the role of a pairing potential. Almost ten years later, using the same pairing model, P. G. de Gennes was able to obtain a real-space superconductivity's theory, nowadays known as Bogolyubov-deGennes formulation\cite{degennesBook}, able to describe spatially non-homogeneous systems. Bogolyubov-deGennes's theory significantly influenced the field of mesoscopic superconductivity\cite{btk} because, in this context, the presence of interfaces disrupts translational invariance, rendering the original BCS theory inadequate. Bogolyubov-deGennes's formulation assumes, according to Gor'kov\cite{fetter}, that fermions are paired by the contact interaction $H_P=-V\int \Psi^{\dag}_{\uparrow}(r) \Psi^{\dag}_{\downarrow}(r) \Psi_{\downarrow}(r) \Psi_{\uparrow}(r) dr$, which is treated by means of a self-consistent mean-field approach similar to the Hartree-Fock method. The pairing Hamiltonian, however, can also be presented in the form $H_P=-V \int n_{\uparrow}(r)n_{\downarrow}(r) dr$ with $n_{\sigma}(r)=\Psi^{\dag}_{\sigma}(r) \Psi_{\sigma}(r)$, which can be recognized as an attractive Hubbard model. The difference between the attractive Hubbard model and the BCS pairing is evident when they are compared in momentum space, so that one is legitimated to ask about the full equivalence between the two pairing models.\\
In a pioneering work\cite{pensonkolb}, Penson and Kolb discussed some of the inadequacies of the attractive Hubbard model in mimicking the superconducting interaction in real space. Following Penson and Kolb, one of the main difficulties is that, within the infinite interaction limit, any Hubbard-type model becomes classical, which raises questions about its potential applications in superconductivity. Thus, inspired by a work of Kulik and Pedan\cite{kulik}, these authors have introduced a real-space pairing mechanism based on nearest neighbors pair-hopping processes. The Penson-Kolb model favors the formation of pairs also promoting their mobility, being this aspect profoundly different from the aforementioned attractive Hubbard model.\\
It is rather astonishing that, as long as a one-dimensional translational-invariant system is considered, an infinite-range Penson-Kolb pairing mechanism coexisting with an attractive Hubbard term is exactly equivalent to the BCS pairing term, i.e. the Fourier transform of the latter provides the former. The formal proof of this statement has been provided in Ref. [\onlinecite{rBCS}] along with relevant observations about the Richardson's integrability of the model. The pairing mechanism proposed in Ref. [\onlinecite{rBCS}] as a real-space BCS model, and here recognized as an infinite-range Penson-Kolb interaction distorted by a local Hubbard term, boasts a number of interesting properties. In particular, one can shows the Richardson's integrability\cite{richardson1,richardson2} of the model under rather generic conditions including onsite potential and hopping disorder, arbitrary lattice structures and values of the interaction strength. All the aforementioned conditions are extremely interesting in order to investigate the interplay between coherence and correlation effects, non-trivial lattice structures and disorder.\\
In this work, we introduce a non-trivial generalization of the real-space BCS model, incorporating an electromagnetic phase. The resulting model is particularly significant as it facilitates analysis of the system's reaction to external stimuli, thereby unlocking insights into its transport properties. Our methodology integrates exact diagonalization, density matrix renormalization group (DMRG) techniques, and the numerical solution of Richardson's equations. This thorough approach enables us to examine critical metrics, such as the system's conductivity in relation to interaction strength. Consequently, we explore the BCS-BEC crossover in a seamless manner, illuminating key aspects of superconducting pairing. Our results offer potential for experimental validation in condensed matter physics, or with heightened precision, through \textit{atomtronics} platforms\cite{atomtronics}.\\
The work is organized as follows. In Section \ref{sec:qrmodel}, we introduce the Hamiltonian model describing fermions confined within a ring-shaped lattice, subject to the simultaneous influence of an Ahronov-Bohm electromagnetic phase and a pairing interaction. We specifically focus on a real-space BCS pairing model, observing that it can be understood as an infinite-range Penson-Kolb pairing mechanism coexisting with an attractive Hubbard term. Furthermore, we compare the real-space BCS model with the attractive Hubbard interaction. In Section \ref{sec:fluxmbgs}, we delve into the generalities concerning the flux-dependence of the ground state energy of a many-body system. We explore various physical observables and their relationships with energy-flux curves. Moving on to Section \ref{sec:enNI}, we present the magnetic response of a many-body system in the absence of a pairing interaction. This section, designed as a reference for subsequent discussions, introduces the energy-flux curves and their sensitivity to the parity of the particle number. In Section \ref{sec:twobp}, we delve into the two-body problem, discussing analogies and peculiarities of the distinct pairing models analyzed. In Section \ref{sec:few}, we provide a thorough examination of the few-body problem, highlighting the Richardson's integrability of the real-space BCS model under appropriate flux bias, backed by extensive DMRG simulations. In this context, features reminiscent of the BCS-BEC transition are also discussed. Finally, conclusions and future perspectives are presented in Section \ref{sec:concl}. Computational details are provided in Appendix \ref{app:A}, while the scaling behavior of charge stiffness with system size is discussed in Appendix \ref{app:B}.

\section{The model}
\label{sec:qrmodel}
Let us consider fermions confined in a ring and subject to the concomitant action of a pairing interaction and the electromagnetic phase induced by a non-vanishing vector potential $\vec{A}$. This condition can be studied either in a condensed matter context and in atomtronics platforms, where artificial gauge fields can be obtained, for instance, by optical imprinting of a phase gradient\cite{roatiImprint}. The kinetic part $H_R$ of the Hamiltonian describing this system can be written in terms of fermionic creation/annihilation operators $a^{\dag}_{\ell\sigma}/a_{\ell\sigma}$, so that
\begin{eqnarray}
\label{eq:ringHam}
H_R=-t \sum_{\ell=1,\sigma}^{N-1} a^{\dag}_{\ell+1\sigma}a_{\ell\sigma}-t \sum_{\sigma}e^{-i 2\pi f}a^{\dag}_{1\sigma}a_{N\sigma}+h.c.,
\end{eqnarray}
while $t$ represents the hopping integral, $f=\Phi/\Phi_0$ stands for the magnetic flux $\Phi=\oint\vec{A} \cdot d\vec{l}$ normalized to the magnetic flux quantum $\Phi_{0}=h/e$, $\sigma \in \{\uparrow,\downarrow\}$ specifies the particle's spin projection and $\ell \in \{1,...,N\}$ is a site index bounded from above by the number $N$ of lattice sites. The interaction part of the Hamiltonian, describing the tendency to form fermions pairs, takes the following form:
\begin{eqnarray}
\label{eq:interactionHam}
H_I=-g \sum_{\ell, r}\Gamma_{\ell r} a_{\ell\uparrow}^{\dag}a_{\ell\downarrow}^{\dag}a_{r\downarrow}a_{r\uparrow},
\end{eqnarray}
where $g>0$ is the pairing strength, while $\Gamma_{\ell r}$ defines the pairing mechanism and the interaction range. In particular, by setting $\Gamma_{\ell r}=\delta_{\ell r}$, an attractive Hubbard interaction is obtained, while, considering $\Gamma_{\ell r}=1/N$, we end up with the real-space BCS interaction introduced in Ref. [\onlinecite{rBCS}]. Remarkably, the real-space BCS interaction can also be understood as an infinite-range Penson-Kolb pairing term \cite{pensonkolb}, which is further modulated by an attractive Hubbard interaction\cite{nota1}. Moreover, the standard Penson-Kolb pairing Hamiltonian, which features a nearest-neighbor pair-hopping term, can be derived through a suitable choice of $\Gamma_{\ell r}$. Therefore, Eq. (\ref{eq:interactionHam}) seems sufficiently versatile to encompass any s-wave pairing model.\\
It is rather instructive expressing the full Hamiltonian $H=H_R+H_I$ in terms of new fermionic fields, i.e. $c_{r\sigma}$, according to the unitary transformation $a_{r\sigma}=e^{i r \varphi}c_{r\sigma}$, with $\varphi=2\pi f/N$. Implementing the aforementioned transformation, the kinetic and the interaction part of the transformed Hamiltonian $\mathcal{H}=\mathcal{H}_R+\mathcal{H}_I$, namely $\mathcal{H}_R$ and $\mathcal{H}_I$, read:
\begin{eqnarray}
&&\mathcal{H}_R=-t \sum_{\ell=1,\sigma}^{N-1} e^{-i \varphi}c^{\dag}_{\ell+1\sigma}c_{\ell\sigma}-t \sum_{\sigma}e^{-i \varphi}c^{\dag}_{1\sigma}c_{N\sigma}+h.c. \nonumber\\
&&\mathcal{H}_I=-g \sum_{\ell, r}\Gamma_{\ell r}e^{-2i(\ell-r)\varphi} c_{\ell\uparrow}^{\dag}c_{\ell\downarrow}^{\dag}c_{r\downarrow}c_{r\uparrow},
\end{eqnarray}
so that the total electromagnetic phase $2\pi f$ results uniformly distributed along the entire ring, being the latter condition reminiscent of the familiar outcome of the Peierls substitution. More surprisingly, a closer inspection to the interaction term $\mathcal{H}_I$ reveals the presence of the electromagnetic phase factor $e^{-2i(\ell-r)\varphi}$. The latter, however, disappears when the Hubbard interaction is considered (i.e., by setting  $\Gamma_{\ell r}=\delta_{\ell r}$), so that the resulting model collapses into the Fermi-Hubbard Hamiltonian of a quantum ring previously considered in literature\cite{minguzziRing}. On the other hand, the phase factor survives when the real space BCS pairing is considered ($\Gamma_{\ell r}=1/N$) because of the nonlocal nature of this type of interaction. In order to understand the role of the unusual phase factor in front of the pairing term, we observe that the pairing Hamiltonian $\mathcal{H}_I$ with $\Gamma_{\ell r}=1/N$ can be understood as a nonlocal hopping from the lattice position $r$ to $\ell$ of paired fermions with opposite spin projection, so that, sometimes, this kind of term is called pair-hopping Hamiltonian\cite{dolcini1,dolcini2}. In view of this observation and in accordance with Ref. [\onlinecite{pensonkolbphase}], the phase factor $e^{-2i(\ell-r)\varphi}$ can be interpreted as the electromagnetic phase acquired by a fermions pair in the nonlocal \textit{hopping} process. The presence of this phase factor is crucial and indeed, as long as a real space BCS interaction is considered, it is possible to verify that omitting it leads to an Hamiltonian model whose ground state energy presents an anomalous dependence upon the applied flux $f$. In particular, numerical analyses show that such anomalous behavior is not compatible with the magnetic response of a superconduting ring.\\
Interestingly, there exists a more subtle argument supporting the necessity to include the phase factor $e^{-2i(\ell-r)\varphi}$ in $\mathcal{H}_I$, when the real space BCS pairing is considered.\\
In order to present the mentioned argument, let us write $\mathcal{H}$ in momentum space. The transformation is implemented by using the annihilation operators expansion $c_{\ell \sigma}=\sum_q \phi_{q}(\ell)c_{q \sigma}$ written in terms of the eigenfunctions $\phi_{q}(\ell)=e^{iq\ell}/\sqrt{N}$ of the noninteracting problem (i.e. the problem obtained by setting $g=0$ in $\mathcal{H}$). Approaching the thermodynamic limit, such wavefunctions are labelled by quantized, but rather dense, values of linear momentum $q=2\pi s/N$, with $s \in \{1,...,N\}$. Thus, the momentum space Hamiltonian can be written as:
\begin{eqnarray}
\label{eq:momHam}
\mathcal{H}=\sum_{q \sigma}E_{q}(\varphi) c^{\dag}_{q\sigma}c_{q\sigma}-\frac{g}{N}\sum_{qp}c_{q\uparrow}^{\dag}c_{\widetilde{q}\downarrow}^{\dag}c_{\widetilde{p}\downarrow}c_{p\uparrow},
\end{eqnarray}
where $E_{q}(\varphi)=-2t \cos(q+\varphi)$ is the flux-dependent single-particle energy spectrum, while the quantities $\widetilde{q}=2\pi-q-2\varphi$ and $\widetilde{p}=2\pi-p-2\varphi$ have been introduced. In the absence of electromagnetic flux (i.e., setting $\varphi=0$ in $\mathcal{H}$), the interaction part of the Hamiltonian promotes the formation of pairs of fermions with opposite spin projection and linear momentum. Such single-particle states, labelled by linear momenta $q$ and $2\pi-q$, are degenerate in energy (i.e., $E_{q}(0)=E_{2\pi-q}(0)$) and the resulting model describes fermions affected by the usual s-wave BCS pairing. Thus, the paired state presents vanishing center of mass momentum, i.e. $q+(2\pi-q)=2\pi$.\\
On the other hand, considering $\varphi \neq 0$ in $\mathcal{H}$, one observes that single-particle states with linear momenta $q$ and $\widetilde{q}=2\pi-q-2\varphi$ are paired by the interaction. These single-particle states are energy degenerate (i.e., $E_{q}(\varphi)=E_{\widetilde{q}}(\varphi)$), while one can explicitly observe that the degeneracy of states labeled by $q$ and $2\pi-q$ is removed by the magnetic flux, so that $E_{q}(\varphi)\neq E_{2\pi-q}(\varphi)$. The paired state $(q\uparrow,\widetilde{q}\downarrow)$ has a net momentum $q+\widetilde{q}=2\pi-2\varphi$, which is induced by the magnetic response of the system. Moreover, the net momentum is the same for all pairs, signaling the presence of a net current flow sustained by paired fermions. The latter observation is suggestive of the formation of persistent currents, which are an important part of the magnetic response of a superconducting system.\\
Once the model's features have been explained, it is rather instructive to observe that, as long as the real space BCS pairing is considered, the presence of the electromagnetic phase factor $e^{-2i(\ell-r)\varphi}$ in $\mathcal{H}_I$ guarantees that: (i) energy degenerate single-particle states are coupled by the interaction term; (ii) the center of mass momentum of a fermions pair is different from zero when $\varphi \neq 0$, it is dependent upon the magnetic flux and identical for all pairs in the system.\\
Conversely, neglecting the phase factor $e^{-2i(\ell-r)\varphi}$ in $\mathcal{H}_I$ induces the formation of fermion pairs originating from single-particle states labelled by $q\uparrow$ and $2\pi-q \downarrow$, which no longer exhibit energy degeneracy for non-vanishing magnetic flux. The coupling of non-degenerate single-particle states represents a costly pairing mechanism and, indeed, it appears unappropriate for the description of conventional superconductors.

\section{Flux-dependence of the many-body ground state energy}
\label{sec:fluxmbgs}
Hereafter, we study the flux-dependent behavior of the ground state energy ($E_G$) of a many-body system described by the Hamiltonian model outlined in Sec. \ref{sec:qrmodel}. The motivation behind this investigation stems from the recognition that the $E_G$ \textit{vs} $f$ curves contain valuable insights into the characteristics of the many-body system. In particular, the flux periodicity of $E_G$ yields crucial information regarding the electric charge of the carriers, while the derivative of $E_G$ with respect to $f$ provides information about the zero-temperature persistent current ($I_{PC}$), in accordance to the relation:
\begin{eqnarray}
\label{eq:pc}
I_{PC} = -\frac{1}{\Phi_0} \frac{dE_G}{df}.
\end{eqnarray}
Interestingly, persistent currents are measurable in a broad spectrum of systems, ranging from condensed matter to \textit{atomtronics} platforms. Consequently, in these systems, the relevant insights contained in $E_G$ can be extracted through direct measurement of $I_{PC}$. In particular, the flux dependence of the persistent current $I_{PC}$ allows to extract the so-called charge stiffness.\\
Charge stiffness constant
\begin{eqnarray}
\label{eq:stiff}
D_c=\frac{N}{\Phi_0^2}\Bigl(\frac{1}{2}\frac{d^2 E_G}{df^2}\Bigl)_{f=0},
\end{eqnarray}
first discussed by W. Kohn\cite{kohn} in the context of the metal-insulator transition, can be used to characterize the electronic conductivity of a system because it is conceptually equivalent to the Drude weight in the optical conductivity and takes vanishing values in insulators.\\
Moreover, in mesoscopic systems, which are the object of the present analysis, $E_G$ also contains precious information about finite-size effects and can exhibit sensitivity to the \textit{parity} of the particles' number. Both of these effects allow to extract information about important length scales related to the many-body problem and gain insights into the quantum statistics of composite particles stabilized by the interaction.

\section{Properties of $E_G$ in the non-interacting case ($g=0$)}
\label{sec:enNI}
This section serve as a review of established results and it is written in order to fix the notation and facilitate a comparison with the interacting case. Thus, we discuss the properties of the ground state energy for a system of free fermions, whose dynamics is determined by the kinetic part of the Hamiltonian presented in Eq. (\ref{eq:momHam}). In particular, we assume that an even number $M \leq 2N$ of fermions is distributed on a $N$-sites quantum ring pierced by a magnetic flux. As anticipated before, the single-particle problem admits the energy spectrum $E_{s}(f)=-2t \cos[\frac{2\pi}{N}(s+f)]$ with $s \in \{1,...,N\}$. Hence, the many-body ground state is achieved by populating single-particle levels with the objective of minimizing the system's total energy while adhering to Pauli's principle. According to this prescription, the ground state energy can be written as $E_G(f)=2 \sum_{n=1}^{M/2}\mathcal{S}_{n}$, where $\mathcal{S}_n$ represents the nth element of the list $\mathcal{S}$, which is obtained by arranging in ascending order the single-particle energy spectrum $\{E_{1}(f),..., E_{N}(f)\}$.\\
\begin{figure}
\includegraphics[scale=1.2]{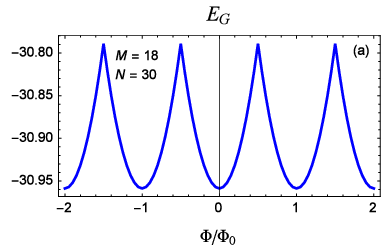}\\
\vspace{5mm}
\includegraphics[scale=1.2]{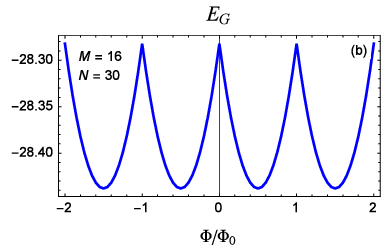}
\caption{Ground state energy $E_G$ (measured in units of the hopping parameter $t$) of a free fermions system as a function of the normalized applied flux $f=\Phi/\Phi_0$. The ring contains $N=30$ lattice sites, while the particles number has been fixed to $M=18$ (panel (a)) or $M=16$ (panel (b)). Flux-one periodicity of the $E_G$ vs $f$ curves is evident for both the panels, so that $E_G(f+1)=E_G(f)$. Moreover, depending on the odd or even parity of $M/2$, the $E_G$ vs $f$ curves show a local minimum or a local maximum, respectively, in close vicinity of $f=0$.}
\label{figure1}
\end{figure}
In Fig. \ref{figure1}, we present the main features of the ground state energy $E_G$ of a free fermions system as a function of the normalized applied flux $f=\Phi/\Phi_0$. In this figure, as in others throughout this article, quantities with dimensions of energy such as the ground state energy $E_G$ or the interaction strength $g$ will be expressed in units of the hopping parameter $t$. In order to be definite, we study a quantum ring containing $N=30$ lattice sites, while fixing the particles number to $M=18$ (panel (a)) or $M=16$ (panel (b)). Flux-one periodicity of the $E_G$ vs $f$ curves is clearly seen for both the panels, so that $E_G(f+1)=E_G(f)$. Moreover, depending on the odd or even parity of $M/2$, the $E_G$ vs $f$ curves show a local minimum or a local maximum, respectively, in close vicinity of $f=0$. A systematic numerical analysis of the $E_G$ vs $f$ curves, exploring different system sizes $N$ and fillings $M$, shows a sort of universal behavior of the quantity $E_G(f)/E_G(0)$, with $E_G(0)$ the ground state energy in the absence of flux. In particular, one observes that $E_G(f) = E_G(0)Z_{\mu}(f)$, with $Z_{\mu}(f)$ exhibiting sensitivity to the parity of $M/2$. Consequently, the dichotomic index $\mu$, belonging to the set $\{o, e\}$, assumes the value $o$ in the case of odd parity and $e$ in the case of even parity. Moreover, $Z_{\mu}(f)$ takes the following form:
\begin{eqnarray}
Z_{\mu}(f)=\frac{\cos \Bigl [\frac{2 \pi}{N}\Omega^{(\mu)}(f) \Bigl ]}{\cos \Bigl [\frac{2 \pi}{N}\Omega^{(\mu)}(0) \Bigl ]},
\end{eqnarray}
where $\Omega^{(\mu)}(f)=\|f+\delta_{\mu,e}/2\|-f-\delta_{\mu,e}/2$ is written in terms of the Kronecker delta function $\delta_{\mu,\mu'}$, while $\|f\|$ stands for the integer number closest to $f$. Since the flux-one periodicity of the $E_G(f)$ curves, and hence of $Z_{\mu}(f)$, is related to the elementary charge of the carriers (i.e. unpaired fermions), it is relevant investigating the harmonic content of the $Z_{\mu}(f)$ curves. In view of the symmetry $Z_{\mu}(-f)=Z_{\mu}(f)$, the Fourier series can be presented in the following form:
\begin{eqnarray}
Z_{\mu}(f)=\sum_{n=0}^{\infty}c_{n}^{(\mu)} \cos(2 \pi n f ),
\end{eqnarray}
where $c_1^{(\mu)}$ contains information about the flux-one periodicity of the curves. Moreover, in virtue of the relation $Z_{e}(f)=Z_{o}(f+1/2)/\cos(\pi/N)$, it is straightforward to demonstrate the validity of the following relation:
\begin{eqnarray}
c_n^{(e)}=\frac{c_n^{(o)}(-)^n}{\cos(\frac{\pi}{N})},
\end{eqnarray}
which immediately implies $|c_n^{(e)}|\approx |c_n^{(o)}|$ when the thermodynamic limit is approached (i.e., for $N \gg 1$). Numerical analysis agrees with the above expectations so that we only present the harmonic content of $Z_o(f)$, which is shown in Fig.\ref{figure2}(a). In particular, in Fig. \ref{figure2}(b), we present the dependence of the coefficients $|c_{n}^{(o)}|$ as a function of $n>0$, which is a relevant figure of merit of the magnetic response of the system. A close inspection of Fig. \ref{figure2}(b) shows the important role of $c_1^{(o)}$, whose amplitude is higher than the second harmonic contribution $c_2^{(o)}$, a rather expected feature for a system of unpaired fermions.\\
A noteworthy observation, particularly in light of the subsequent analysis, pertains to the relative amplitudes of the coefficients $c_1^{(o)}$ and $c_2^{(o)}$. The intensity of these coefficients plays a crucial role in determining the dependence of the ground state energy on the applied magnetic field. Specifically, $c_1^{(o)}$ introduces a periodicity of $\Phi_0=h/e$, while $c_2^{(o)}$ contributes to a period halving, corresponding to a $\Phi_0/2$ periodicity. Recent literature\cite{barash2008} underscores the nontrivial interplay between the $\Phi_0$ and $\Phi_0/2$ periodicity in $E_G(f)$. Indeed, it has been demonstrated that an $h/e$ periodicity can also manifest under appropriate conditions in a system of fermions subjected to the BCS pairing interaction.\\
In the following, we will demonstrate that some of the conclusions reached in Ref. [\onlinecite{barash2008}] within the framework of a mean-field treatment of the BCS interaction retain their validity also when the mean-field approximation is removed and a genuine interacting many-body system is considered.

\begin{figure*}
\includegraphics[scale=1.4]{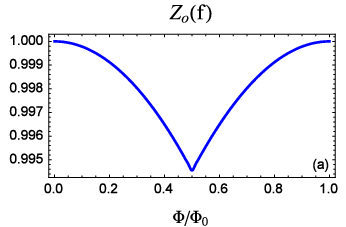}
\includegraphics[scale=1.4]{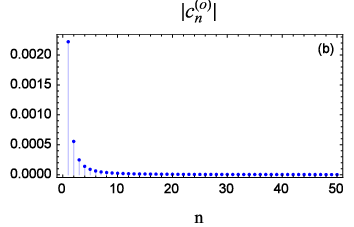}
\caption{Panel (a): $Z_{o}(f)$ curve in the domain $f \in [0,1]$. Panel (b): Absolute value of the Fourier coefficients $c_n^{(o)}$, with $n>0$, of the curve shown in panel (a).}
\label{figure2}
\end{figure*}

\section{Two-body problem}
\label{sec:twobp}
This section addresses the relevant problem of two interacting fermions confined to a quantum ring subject to the perturbing effect of a magnetic flux. The two-body problem can be easily addressed by using the Hamiltonian $H=H_R+H_I$, with $H_R$ and $H_I$ given in Eq. (\ref{eq:ringHam}) and (\ref{eq:interactionHam}), respectively. Moreover, $H$ can be rearranged in the compact form:
\begin{eqnarray}
\label{eq:2bodyHam}
H=\sum_{\ell r \sigma}a^{\dag}_{\ell\sigma}\mathbb{H}_{\ell r}^{(f)} a_{r\sigma}-g \sum_{\ell, r}\Gamma_{\ell r} a_{\ell\uparrow}^{\dag}a_{\ell\downarrow}^{\dag}a_{r\downarrow}a_{r\uparrow},
\end{eqnarray}
where $\mathbb{H}_{\ell r}^{(f)}$ is an appropriate flux-dependent hopping matrix. The two-body problem is conveniently treated adopting a first quantization formalism. Thus, we introduce a general two-particle state:
\begin{eqnarray}
|\Psi\rangle=\frac{1}{2}\sum_{x_1\sigma_1, x_2 \sigma_2}\psi_{\sigma_1 \sigma_2}(x_1,x_2)|x_1 \sigma_1, x_2 \sigma_2\rangle,
\end{eqnarray}
which is superposition of Slater determinants $|x_1 \sigma_1, x_2 \sigma_2\rangle=a_{x_2 \sigma_2}^{\dag}a_{x_1 \sigma_1}^{\dag}|0\rangle$ describing two particles with spin projection $\sigma_1$ and $\sigma_2$ and located at lattice positions $x_1$ and $x_2$, respectively. Within this framework, $\psi_{\sigma_1 \sigma_2}(x_1,x_2)$ is the first quantization wavefunction, while $|0\rangle$ represents the empty lattice state. The requirement that $|\Psi\rangle$ is an eigenstate of the Hamiltonian $H$ in Eq. (\ref{eq:2bodyHam}) allows to write the stationary Schr\"{o}dinger equation $H |\Psi\rangle=E |\Psi\rangle$, being $E$ the energy eigenvalue. Projecting the Schr\"{o}dinger equation on a single Slater determinant, namely $|y_1 s_1, y_2 s_2\rangle$, one obtains:
\begin{eqnarray}
\label{eq:schrodingerProj}
\langle y_1 s_1, y_2 s_2|H_R+H_I|\Psi\rangle=E \ \psi_{s_1 s_2}(y_1,y_2).
\end{eqnarray}
On the other hand, direct computation shows that:
\begin{eqnarray}
\label{eq:expHr}
&&\langle y_1 s_1, y_2 s_2|H_R|\Psi\rangle=\nonumber\\
&=&\sum_{y} \Bigl ( \mathbb{H}_{y_2 y}^{(f)} \psi_{s_1s_2}(y_1,y)+\mathbb{H}_{y_1 y}^{(f)} \psi_{s_1s_2}(y,y_2) \Bigl ),
\end{eqnarray}
being this term related with the two-particle hopping on the ring. When the interaction part of the Hamiltonian is considered, distinct results are obtained depending on the considered pairing model. In general, the interaction only couples particles with opposite spin projection ($s_1 \neq s_2$) in a spin singlet state so that, hereafter, we focus on this relevant case. In particular, when the attractive Hubbard model is considered, one obtains:
\begin{eqnarray}
\label{eq:expHiHubbard}
\langle y_1 s_1, y_2 s_2|H_I|\Psi\rangle=-g \delta_{y_1 y_2} \psi_{s_1 s_2}(y_1,y_2),
\end{eqnarray}
while a rather different result is derived for the real space BCS pairing, i.e.,
\begin{eqnarray}
\label{eq:expHiBCS}
\langle y_1 s_1, y_2 s_2|H_I|\Psi\rangle=-\frac{g}{N} \delta_{y_1 y_2} \sum_y \psi_{s_1 s_2}(y,y).
\end{eqnarray}
Equation (\ref{eq:schrodingerProj}), complemented by Eqs. (\ref{eq:expHr})-(\ref{eq:expHiHubbard}) or (\ref{eq:expHr}) and (\ref{eq:expHiBCS}), univocally defines the stationary Schr\"{o}dinger equation for the two-body wavefunction $\psi_{s_1 s_2}(y_1,y_2)$ with $s_1 \neq s_2$. Moreover, since $H=H_R+H_I$ preserves the total number of particles $\hat{N}$ and the number $\hat{N}_{\sigma}$ of particles with spin projection $\sigma \in \{\uparrow,\downarrow\}$, the two-particle problem described by $\psi_{s_1 s_2}(y_1,y_2)$ with $s_1 \neq s_2$ is not coupled with the equal spin problem ($s_1=s_2$) because different spin sectors are independent. Furthermore, since fermions are in a singlet state, the Schr\"{o}dinger equation in (\ref{eq:schrodingerProj}) has to be solved under the requirement that $\psi_{s_1 s_2}(y_1,y_2)=\psi_{s_1 s_2}(y_2,y_1)$, imposed by the global antisymmetry of the two-body wavefunction under particle exchange.\\
Once the Schr\"{o}dinger equation has been numerically solved as detailed in Ref. [\onlinecite{rBCS}], it is possible to study the flux dependence of the energy spectrum, which defines the magnetic response of the system and provides information about the typical size of a two-particle bound state (i.e., a fermions pair). Since the above features are expected to depend on the specific form of the pairing interaction, in the following we discuss the peculiarities of the distinct pairing models presented in Eqs. (\ref{eq:expHiHubbard}) and (\ref{eq:expHiBCS}), respectively. In comparison to a similar analysis documented in Ref. [\onlinecite{rBCS}], the impact of the magnetic flux on the system in this study enables us to emphasize additional similarities and distinctions among the contemplated pairing mechanisms.
\begin{figure*}[!t]
\includegraphics[scale=1.2]{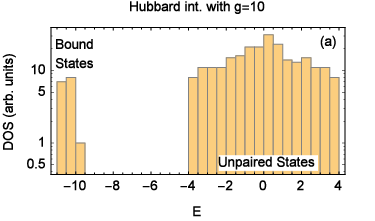}
\includegraphics[scale=1.2]{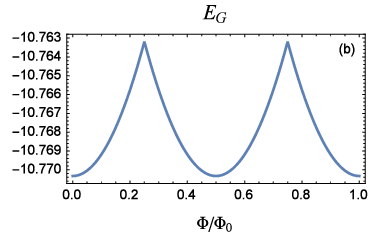}
\includegraphics[scale=1.2]{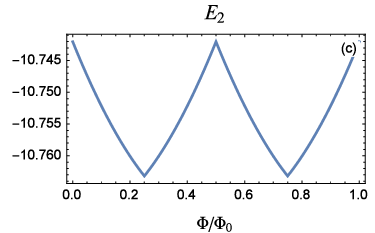}
\includegraphics[scale=1.2]{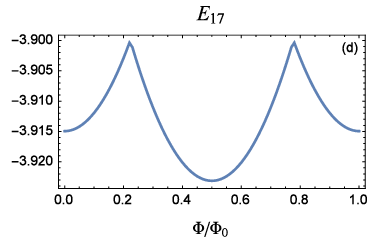}
\includegraphics[scale=1.09]{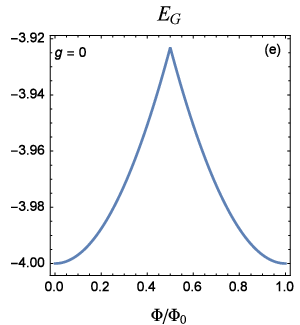}
\includegraphics[scale=1.13]{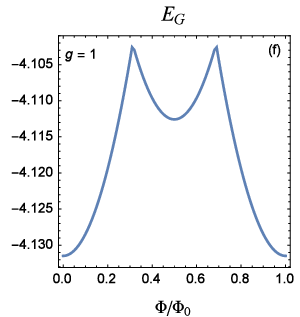}
\includegraphics[scale=1.13]{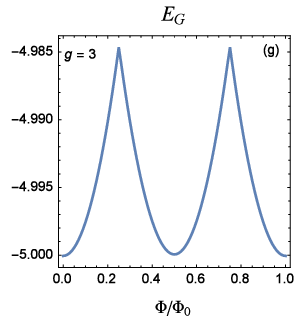}
\caption{Panel (a): Density of states (DOS) of the problem of two fermions subject to attractive Hubbard interaction obtained by setting the interaction strength to $g=10$, the number of lattice sites to $N=16$, while fixing $f=0$. The energy spectrum is organized in two distinct parts: (i) The bound states region formed by fermions paired by the interaction; (ii) The unpaired states region, describing free fermions. The energy required to produce unpaired fermions starting from a fermions pair is controlled by the interaction strength $g$. Looking at the flux dependence of the energy levels, one observes that bound states present a $\Phi_0/2$ periodicity, which is related with the formation of a fermions pair. Unpaired states, on the other hand, are characterized by $\Phi_0$ periodicity of the energy-flux curves, reminiscent of the free fermions physics. The energy-flux curves for the ground state, the first excited state and the bottom of the band formed by the unpaired states are presented in panels (b), (c) and (d), respectively. Panels (e), (f), (g) show the energy-flux curves for the ground state of the two-fermions problem as a function of the interaction strength $g$, while setting $N=16$. It is shown that the flux periodicity of the energy-flux curves evolves from $\Phi_0$ towards $\Phi_0/2$ as the interaction strength increases. A similar behavior is observed for fixed interaction strength $g$ and variable system size $N$.}
\label{figure3}
\end{figure*}
In Fig. \ref{figure3}, we provide a bird's-eye view of the main features of the problem of two particles subject to attractive Hubbard interaction. In Fig. \ref{figure3}(a), in particular, we depict the density of states of the two-particle energy spectrum for a system of $N=16$ lattice sites and in the absence of magnetic flux ($f=0$), while, for presentation reasons, a rather intense interaction strength ($g=10$) has been used. The energy spectrum is organized into two parts with distinct properties: (i) The region of bound states, wherein fermions form pairs stabilized by the interaction; (ii) the region of unpaired states, representing free fermions. The energy needed to generate unpaired fermions from a fermion pair is governed by the interaction strength $g$. Furthermore, differently from the behavior of a Cooper's pair \cite{cooper56}, the elementary excitation of a pair corresponds to increasing its kinetic energy rather than to the splitting of the pair. Examining the flux dependence of the energy levels reveals that bound states exhibit a periodicity of $\Phi_0/2$ associated with fermion pair formation. In contrast, unpaired states display an energy-flux curve with a periodicity of $\Phi_0$, reminiscent of free fermion physics. Panels (b), (c), and (d) support the above statements and illustrate the energy-flux curves for the ground state ($E_G$), the first excited state ($E_{2}$), and the bottom of the band formed by unpaired states ($E_{17}$), respectively. Additionally, panels (e), (f), and (g) showcase the energy-flux curves for the ground state of the two-fermion problem for different values of the interaction strength $g$, while keeping $N=16$. Notably, these curves demonstrate a transition in flux periodicity from $\Phi_0$ to $\Phi_0/2$ as the interaction strength is increased. A similar behavior is observed by fixing the interaction strength $g$, while letting the system size $N$ increase. These findings agree with the mean-field picture discussed in Ref. [\onlinecite{barash2008}] and suggest that $\Phi_0/2$ periodicity emerges when the system size $N$ is much greater than the typical size $\xi \sim 1/g$ of the bound state formed by paired fermions.\\
A similar analysis can be performed in the case of two particles subject to the real-space BCS interaction. In Fig. \ref{figure4}(a), in particular, we present the density of states of the aforementioned problem obtained by setting the interaction strength to $g=10$, the number of lattice sites to $N=16$, while fixing $f=0$. In contrast with the Hubbard case, the energy spectrum presents a single bound state accompanied by a DOS region describing free fermions (unpaired states). The bound state presents a $\Phi_0/2$ periodicity (see Fig. \ref{figure4}(b)), which is related with the formation of a fermions pair. Unpaired states, on the other hand, are characterized by $\Phi_0$ periodicity of the energy-flux curves, reminiscent of the free fermions physics (see Fig. \ref{figure4}(c)). Panels (d), (e), (f) of Fig. \ref{figure4} show the energy-flux curves for the ground state of the two-fermions problem as a function of the interaction strength $g$, while setting $N=16$. Similarly to the Hubbard case and with identical motivation, the flux periodicity of the energy-flux curves evolves from $\Phi_0$ towards $\Phi_0/2$ as the interaction strength increases.\\
In comparison with the attractive Hubbard interaction described in Fig. \ref{figure3}, the main peculiarity of the real-space BCS pairing is the presence of a single bound state, being a pair breaking process the unique way to excite this state. Thus, the single bound state behaves like a fermions pair as described in the seminal paper by Cooper\cite{cooper56}.
\begin{figure*}[!h]
\includegraphics[scale=1.2]{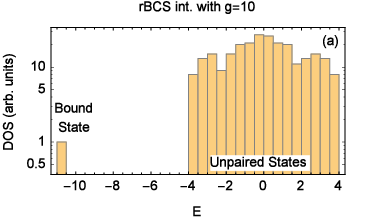}
\includegraphics[scale=1]{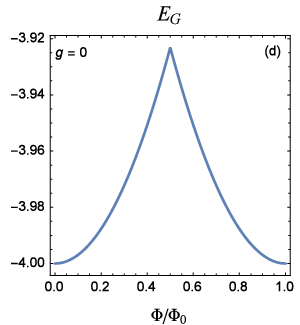}\\
\includegraphics[scale=1.2]{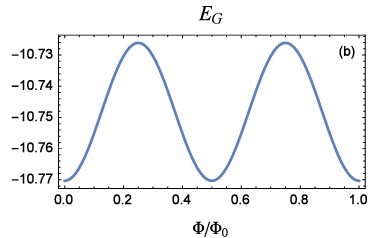}
\includegraphics[scale=1]{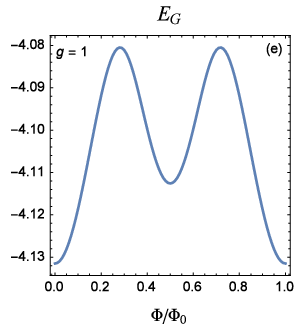}\\
\includegraphics[scale=1.2]{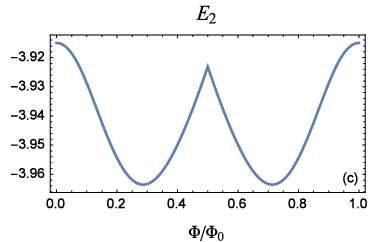}
\includegraphics[scale=1]{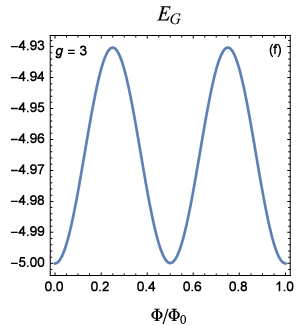}
\caption{Panel (a): Density of states (DOS) of the problem of two fermions subject to a real-space BCS interaction obtained by setting the interaction strength to $g=10$, the number of lattice sites to $N=16$, while fixing $f=0$. In contrast with the Hubbard case, the energy spectrum presents a single bound state accompanied by a DOS region describing free fermions (unpaired states). The energy required to produce unpaired fermions starting from a fermions pair is controlled by the interaction strength $g$. Looking at the flux dependence of the energy levels, one observes that the bound state presents a $\Phi_0/2$ periodicity, which is related with the formation of a fermions pair. Unpaired states, on the other hand, are characterized by $\Phi_0$ periodicity of the energy-flux curves, reminiscent of the free fermions physics. The energy-flux curves for the ground state and the bottom of the band formed by the unpaired states are presented in panels (b) and (c), respectively. Panels (d), (e), (f) show the energy-flux curves for the ground state of the two-fermions problem as a function of the interaction strength $g$, while setting $N=16$. It is shown that the flux periodicity of the energy-flux curves evolves from $\Phi_0$ towards $\Phi_0/2$ as the interaction strength increases. A similar behavior is observed for fixed interaction strength $g$ and variable system size $N$.}
\label{figure4}
\end{figure*}

\subsection{Two-body problem: Charge stiffness and maximum value of the persistent current.}
\label{sec:twobody}
\begin{figure*}[h]
\includegraphics[scale=1.35]{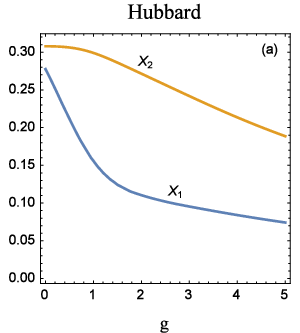}
\includegraphics[scale=1.32]{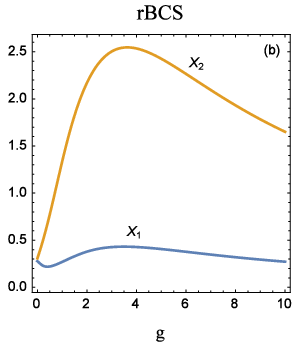}
\caption{$X_{1,2}$ defined in Eqs. (\ref{eq:x1}) and (\ref{eq:x2}) as a function of the pairing strength $g$. The curves have been obtained by solving the two-particle problem with $N=16$, while considering the attractive Hubbard interaction (panel (a)) or the real space BCS pairing (panel (b)). For both the panels, it is observed that $X_1 \propto X_2$ for $g\gtrsim 2$. In particular, taking $g\gtrsim 2$, one obtains $X_1 \approx 0.39 X_2$ for the Hubbard interaction (panel (a)) or $X_1 \approx 0.17 X_2$ for the real space BCS pairing (panel (b)).}
\label{figure5}
\end{figure*}

In order to characterize the transport properties of the system, we introduce the auxiliary functions
\begin{eqnarray}
\label{eq:x1}
X_1 &=& \max_{f \in [0,1]}\frac{dE_G}{df}\\
\label{eq:x2}
X_2 &=& \Bigl(\frac{1}{2}\frac{d^2 E_G}{df^2}\Bigl)_{f=0},
\end{eqnarray}
which are proportional to the maximum value of the persistent current (Eq. \ref{eq:pc}) and to the charge stiffness (Eq. \ref{eq:stiff}), respectively. These quantities are not independent and exhibit proportionality for rather strong interaction values (i.e. $g\gtrsim 2$). On the other hand, for small or moderate values of the pairing constant $g$, the behavior of $X_1$ with respect to $g$ deviates from a simple proportionality relation with $X_2$. This observation can be understood by noticing that the maximum value of the persistent current, and thus $X_1$, is also affected by the flux value $f^{\ast}$ defined as $\max_{f\in[0,1]} I_{PC}(f)=I_{PC}(f^{\ast})$. The value $f^{\ast}$, on its turn, presents a dependence on $g$ which is pronounced only for small or moderate interaction values, while $f^{\ast}$ becomes insensitive to the interaction strength when $g\gtrsim 2$. The dependence of $f^{\ast}$ on $g$ is mainly induced by the evolution of the flux-periodicity of the ground state energy from $\Phi_0$ towards $\Phi_0/2$. When the persistent currents present a sawtooth profile with respect to the applied flux $f$, which is the case for the Hubbard interaction, the flux dependence of the ground state energy within the relevant flux range is well captured by the low-flux expansion $E_G(f) \approx E_G(0)+X_2 f^2$. Thus, we obtain $\Phi_0 I_{PC}(f^{\ast}) \approx -2 X_2 f^{\ast}$ and $\Phi_0 I_{PC}(f^{\ast})=-X_1$, which immediately implies $X_1 \approx 2 X_2 f^{\ast}$. Numerical analysis shows that the argument presented above also works in a quantitative manner for the real space BCS pairing.\\
In Fig. \ref{figure5} we present the behavior of the quantities $X_{1,2}$ as a function of the pairing strength $g$ obtained by setting the system's size to $N=16$. Considering the case of the attractive Hubbard interaction reported in Fig. \ref{figure5}(a), we observe that the charge stiffness, proportional to $X_2$, is a decreasing function of $g$ so that the pairing reduces the system's conductivity. This conclusion is also supported by the analysis of $X_1$ whose behavior indicates a suppression of the persistent currents for increasing values of $g$. Suppression of the persistent currents in the attractive Hubbard model has been reported in literature\cite{giamarchi} and attributed to a velocity, and hence stiffness, reduction when the pairing strength is increased\cite{kawakami}.\\
A rather different behavior is observed for the real space BCS interaction presented in Fig. \ref{figure5}(b). In particular, in that case, one observes an increasing of $X_2$, and thus of the system's conductivity, for $g \lesssim 3$ and a subsequent gentle decrease for interaction values greater than $g \approx 3$. A similar behavior, with some deviations at low $g$ values, is also observed for the $X_1$ \textit{vs} $g$ curve. Interestingly, the $X_2$ \textit{vs} $g$ curve presented in Fig. \ref{figure5}(b) shares similarities with the critical current of a Josephson junction formed by weakly coupled Richardson superconductors\cite{buccheri}. According to Ref. [\onlinecite{buccheri}], the critical current under this condition exhibits a maximum at the BCS-BEC crossover, suggesting an analogous interpretation for the behavior of the $X_2$ \textit{vs} $g$ curve presented in Fig. \ref{figure5}(b).

\subsection{Two-body problem: Size effects and flux periodicity of the ground state energy.}
\begin{figure*}[h]
\includegraphics[scale=1.3]{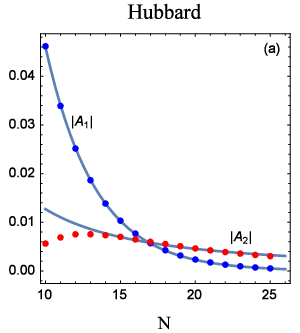}
\includegraphics[scale=1.3]{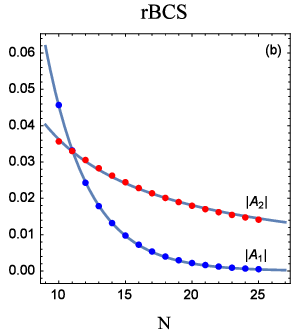}
\includegraphics[scale=1.3]{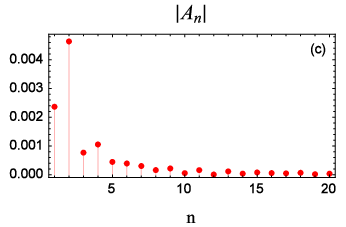}
\includegraphics[scale=1.3]{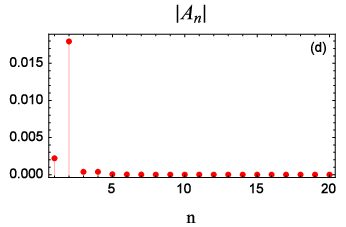}
\caption{Panel (a): Dependence on the system size $N$ of the absolute values of the Fourier coefficients $A_{1,2}$ involved in the ground state energy expansion given in Eq. (\ref{eq:fourierSize}). The numerical analysis has been performed by considering two particles subject to attractive Hubbard interaction with pairing strength given by $g=1.2$. Full lines have been obtained by considering the interpolation curves $|A_{1}| \propto \exp(-N/\lambda)$ and $|A_{2}| \propto N^{\gamma}$, being the interpolation parameters $\lambda \approx 3.34$ and $\gamma \approx -1.48$. Panel (b): The same analysis reported in panel (a) is here specialized to the problem of two particles interacting via the real space BCS interaction with pairing strength given by $g=1.2$. Full lines represent the interpolation curves $|A_{1}| \propto \exp(-N/\lambda)$ and $|A_{2}| \propto N^{\gamma}$, with $\lambda \approx 3.24$ and $\gamma \approx -1.00$. Panel (c): Absolute value of the Fourier coefficients $A_n$ with $n>0$ obtained by setting attractive Hubbard interaction with $g=1.2$ and system size $N=20$. Panel (d): Absolute value of the Fourier coefficients $A_n$ with $n>0$ obtained by setting real space BCS interaction with $g=1.2$ and system size $N=20$.}
\label{figure6}
\end{figure*}

Within the context of the two particle problem, size effects represent another important issue which can be addressed by studying the flux-periodicity of the ground state energy $E_G(f)$. In order to proceed along this line, the ground state energy is expanded in Fourier cosine series,
\begin{eqnarray}
\label{eq:fourierSize}
E_G(f)=\sum_{n} A_n \cos(2\pi n f),
\end{eqnarray}
and the dependence of the coefficients $A_n$ on the system's size $N$ is studied. In order to monitor the $\Phi_0$ to $\Phi_0/2$ transition, we focus on $A_1$ and $A_2$, respectively related to $\Phi_0$ and $\Phi_0/2$ flux-periodicity of $E_G(f)$.\\
The analysis of the size-dependence of the Fourier coefficients $A_{1,2}$ for the problem of two particles paired by the attractive Hubbard interaction is shown in Fig. \ref{figure6}(a). It is shown that the considered coefficients present distinct dependence on the system size $N$. In particular, we find that $A_1$ presents an exponential decay with $N$ to be compared with a slower power-law decay of $A_2$. Due to this difference, when the system exceeds a certain size, $A_2$ becomes greater than $A_1$ and a $\Phi_0/2$ periodicity emerges.\\
A similar behavior is observed when the real space BCS pairing is considered (see Fig. \ref{figure6}(b)). In the latter case, while the behavior of $|A_1|$ remains almost identical to that observed in panel (a), now $|A_2|$ goes to zero as the inverse of the system's size (i.e., $|A_2| \propto N^{-1}$), to be compared to a faster decay of $|A_2|$ observed for the attractive Hubbard interaction.\\
Additional information can be deduced by direct inspection of the full Fourier spectrum of $E_G(f)$. By setting $N=20$ and $g=1.2$, we present the entire Fourier spectrum of $E_G(f)$ for the attractive Hubbard interaction (Fig. \ref{figure6}(c)) and the real space BCS interaction (Fig. \ref{figure6}(d)). While both the Fourier spectra presented in panels (c) and (d) evidence the dominant role of $|A_2|$, harmonics higher than $A_2$ are strongly suppressed only for the real space BCS model, while, considering the Hubbard's case shown in Fig. \ref{figure6}(c), harmonics higher than $A_2$ appear to be still relevant.

\section{Few-body problem}
\label{sec:few}
\begin{figure*}[h]
\includegraphics[scale=0.92]{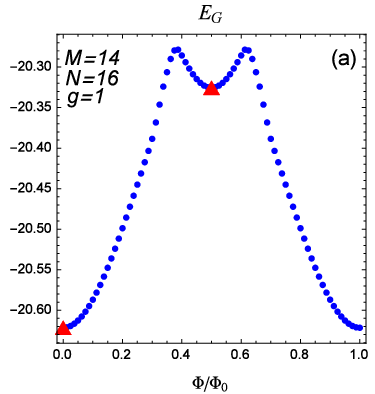}
\includegraphics[scale=0.9]{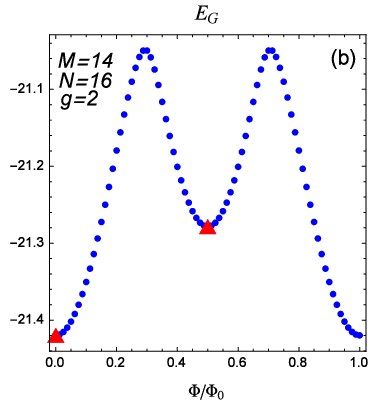}
\includegraphics[scale=0.9]{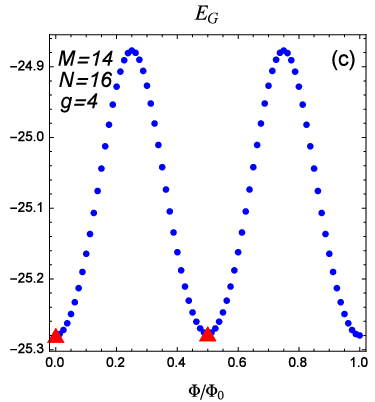}
\includegraphics[scale=0.9]{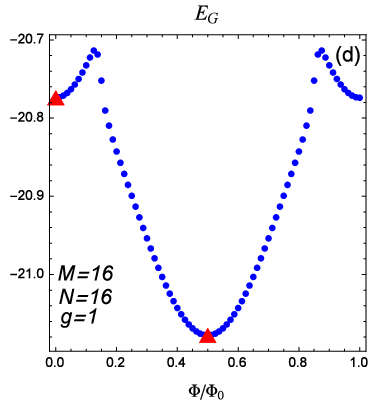}
\includegraphics[scale=0.9]{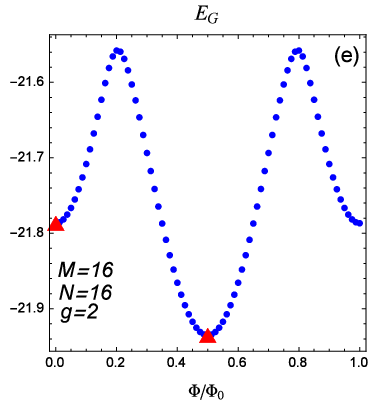}
\includegraphics[scale=0.9]{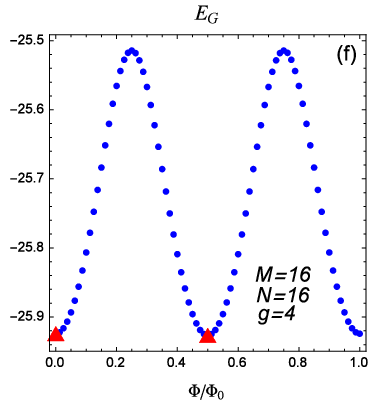}
\includegraphics[scale=1.065]{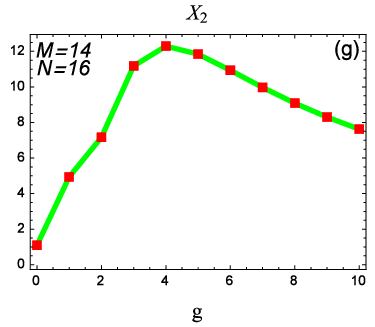}
\includegraphics[scale=1.1]{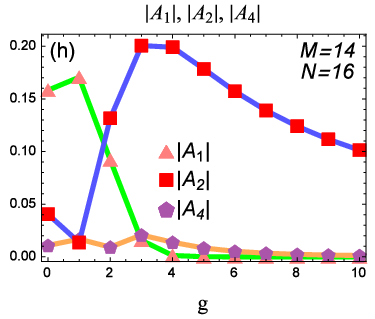}
\caption{Panels (a)-(c): The ground state energy $E_G$ of a fermionic system subject to the real space BCS pairing plotted as a function of the applied magnetic flux $\Phi/\Phi_0$. Blue circular points represent results obtained by using the Density Matrix Renormalization Group technique, while red triangles represent the numerical solution of the Richardson equations. Various curves correspond to systems with $M=14$ particles distributed in a ring with $N=16$ lattice sites and different interaction strengths $g \in \{1,2,4\}$ (refer to the legends for details). Panels (d)-(f): $E_G$ \textit{vs} $\Phi/\Phi_0$ curves obtained by setting $M=16$, while keeping the other parameters unchanged with respect to the first row. Panel (g): $X_2$ \textit{vs} $g$ curve obtained by considering $M=14$ particles and $N=16$ lattice sites. Panel (h): Absolute values of the Fourier coefficients $A_1$, $A_2$, and $A_4$ (see Eq. (\ref{eq:fourierSize})) shown as a function of the interaction strength $g$, while considering $M=14$ particles and $N=16$ lattice sites.}
\label{figure7}
\end{figure*}

So far, we have discussed results pertaining to the two-body problem, highlighting both unique aspects and similarities of the attractive Hubbard interaction and the real-space BCS pairing. The characteristics of the attractive Hubbard interaction are well documented in the many-body physics literature and, for this reason, we present in this section results for few-body systems governed by the real-space BCS pairing.\\
In order to follow our program, let us consider the quantum ring Hamiltonian:
\begin{eqnarray}
\label{eq:fewH}
H &=& -t \sum_{\ell=1,\sigma}^{N-1} a^{\dag}_{\ell+1\sigma}a_{\ell\sigma}-t \sum_{\sigma}e^{-i 2\pi f}a^{\dag}_{1\sigma}a_{N\sigma}+h.c.+\nonumber\\
&-& \frac{g}{N} \sum_{\ell, r} a_{\ell\uparrow}^{\dag}a_{\ell\downarrow}^{\dag}a_{r\downarrow}a_{r\uparrow},
\end{eqnarray}
already discussed before (see Eq. (\ref{eq:ringHam}) and (\ref{eq:interactionHam}) with $\Gamma_{\ell r}=1/N$) and an even number of fermions $M \leq 2N$. Due to the gauge choice, the applied magnetic flux only affects, via the phase factors $e^{\pm i 2\pi f}$, a single hopping linking the lattice sites labelled by $1$ and $N$, respectively. For integer or half integer values of $f$, the Hamiltonian in Eq. (\ref{eq:fewH}) is Richardson integrable because the hopping matrix satisfies the requirements discussed in Ref. [\onlinecite{rBCS}]. In particular, setting $f$ as an integer or a half integer number, the ground state energy of the system in the absence of singly-occupied levels can be presented in the form $E_G=\sum_{\nu \in \{1,...,n_p\}}e_{\nu}$, where the quantities $e_{\nu}$, the so-called \textit{rapidities}, are particular solutions of the coupled algebraic equations:
\begin{eqnarray}
\label{eq:RichardsonEqs}
1+\sum^{n_p}_{\mu =1(\ne \nu)}\frac{2\widetilde{g}}{e_{\mu }-e_{\nu }}=\sum_{j=1}^{N}\frac{\widetilde{g}}{2E_j(f)-e_{\nu }},
\end{eqnarray}
labelled by $\nu \in \{1,...,n_p\}$ and written in terms of the effective interaction strength $\widetilde{g}=g/N$ and of the single-particle energy spectrum $E_j(f)=-2t \cos[\frac{2 \pi}{N}(j+f)]$.\\
All possible solutions of Eq. (\ref{eq:RichardsonEqs}) can be presented in the form $\{e_1^{(\lambda)},...,e_{n_p}^{(\lambda)}\}$, where the maximum value of $\lambda \in \{1,..., \lambda_M \}$ is given by
\begin{eqnarray}
\lambda_M=\frac{N!}{n_p!(N-n_p)!}.
\end{eqnarray}
The latter represents the dimension of the \textit{seniority-zero} Hilbert space, i.e. the number of distinct eigenstates with $n_p=M/2$ pairs in the absence of singly-occupied levels. Although solving Eq. (\ref{eq:RichardsonEqs}) involves non-trivial numerical subtilities\cite{richardson2}, optimization of solution algorithms is possible by observing that generic solutions $e_{\nu}$ are typically found close to $2E_j(f)$ for negligible interaction strength, while they progressively shift towards lower values as the interaction strength increases. Therefore, as long as one is interested in the ground state energy of the interacting system, the energy levels of the filled Fermi sea provide a suitable starting point for algorithms searching for the rapidities $e_{\nu}$.\\
For arbitrary flux values $f$, the usual Richardson procedure, requiring exact energy degeneracy of the paired states, only works in thermodynamic limit. On the other hand, when considering mesoscopic systems the combined effect of finite levels spacing and real values of $f$ does not allow the implementation of the usual Richardson procedure.\\
Thus, in order to present a complete characterization of the magnetic response of the system, we use a density matrix renormalization group (DMRG) technique based on a tensor-network approach \cite{itensor} (see Appendix \ref{app:A} for details). The numerical accuracy of the DMRG simulations was determined by comparing them with Richardson integrable cases. Considering the worst cases (i.e. simulations with $g<2$), discrepancies between the two methods start to manifest from the fourth decimal digit onward, while the accuracy typically improves when simulations with $g>2$ are considered.\\
In Fig. \ref{figure7}, we showcase a DMRG study (blue circular points in panels (a)-(f)) of a system with $N=16$ lattice sites, supplemented by exact results derived from the Richardson integrable cases (red triangles in panels (a)-(f)). In particular, in Fig. \ref{figure7}(a)-(c), we study the flux-dependence of the ground state energy of a fermionic system with $M=14$ particles, a condition in which $M/2$ is an odd number. The $E_G$ \textit{vs} $\Phi/\Phi_0$ curves present a periodic behavior with respect to the applied flux and a periodicity change from $\Phi_0$ (panels (a) and (b)) to $\Phi_0/2$ (panel (c)) as the interaction strength increases. The $\Phi_0$ to $\Phi_0/2$ crossover has been already discussed in the single-pair case and it is rather interesting that the same effect survives, with analogous motivation, even when the many-particle problem is considered. The robustness of the periodicity crossover to different fillings suggests that a fundamental mechanism is working. Indeed, according to Ref. [\onlinecite{combescot2011}], Cooper pairs interact via Pauli blocking only, so that the relevant features of the many-particle ground state are reminiscent of the single-pair physics.\\
To assess the sensitivity of the system to the parity of the particles' number, in Fig. \ref{figure7}(d)-(f), we replicate the analysis conducted in panels (a)-(c) using a system comprising $M=16$ particles, thereby ensuring that $M/2$ is an even number. As already observed in panels (a)-(c), a direct inspection of panels (d)-(f) confirms that an increase of the interaction strength leads to a change in periodicity from $\Phi_0$ (as seen in panels (d) and (e)) to $\Phi_0/2$ (as observed in panel (f)). Interestingly, the parity effect, which is clearly at work when $g\lesssim 2$, is no longer observed when the $\Phi_0/2$ periodicity is established (i.e. for $g\gtrsim 4$). For this reason, the curves in Fig. \ref{figure7}(c) and Fig. \ref{figure7}(f), respectively obtained for odd and even values of $M/2$, tend to coincide. Suppression of the parity effect accompanied by a $\Phi_0/2$ periodicity of the ground-state energy has been reported in Ref. [\onlinecite{minguzziRing}], where the $\Phi_0/2$ periodicity has been commented as a clear signature of the formation of Cooper pairs. The latter conclusion reported in Ref. [\onlinecite{minguzziRing}] has been reached by using a rather elegant argument according to which, assuming strong coupling for both the models, the Fermi-Hubbard Hamiltonian is continuously connected to the Bose-Hubbard one, thus permitting to explore the two sides of the BCS-BEC crossover by means of a piecewise analysis based on the mentioned models.\\
Considering that any purely attractive interaction in reduced spatial dimensions produces a two-particle bound state (and thus a BCS instability)\cite{zwerger}, it is here interesting to complement the suggestive arguments presented in Ref. [\onlinecite{minguzziRing}] with a perhaps equivalent interpretation. According to Ref. [\onlinecite{barash2008}] and as also confirmed by our previous analysis shown in Fig. \ref{figure6}, a $\Phi_0/2$ periodicity can only emerge when the coherence length, i.e. the typical size of a two-particle bound state, is much smaller than the ring circumference, representing the available volume. According to this observation, a $\Phi_0$ to $\Phi_0/2$ crossover can be induced by increasing the system size taking the interaction strength fixed or viceversa. Thus, we arrive at a rather general principle according to which a two-particle bound state induced by attractive interaction behaves as such only if the typical distance among the constituent fermions is limited by the interaction effects rather than by the finiteness of the available volume. This principle can also be understood invoking a relation between quantum confinement and energy quantization. Indeed, the single-particle physics become apparent when the mean level spacing $\lambda$ of the single-particle energy spectrum significantly exceeds the interaction strength $g$. Since $\lambda$ is a decreasing function of the system size $N$, this condition is met in mesoscopic systems, for which quantum confinement effects play a significant role. Conversely, when the limit $\lambda \ll g$ is considered, the many-body physics starts to play a dominant role.\\
Despite certain interpretative subtleties, it is quite intriguing that the physical picture presented in Ref. [\onlinecite{minguzziRing}] is here confirmed by resorting to a completely different Hamiltonian model (Eq. \ref{eq:fewH}), which allows for a seamless exploration of the entire BCS-BEC crossover.\\
The dependence of the ground-state energy on the applied flux provides relevant insights into the system's transport properties. As discussed in Sec. \ref{sec:twobody}, the quantity $X_2$, proportional to the charge stiffness, serves as a measure of the system's conductivity, following W. Kohn's initial proposal (see Appendix \ref{app:B} for details). Accordingly, extensive DMRG simulations have been performed to obtain the $X_2$ \textit{vs} $g$ curve, shown in Fig. \ref{figure7}(g), for a system with $M=14$ particles distributed in a ring of $N=16$ lattice sites. Interestingly, the curve depicted in Fig. \ref{figure7}(g) exhibits qualitative characteristics akin to those seen in Fig. \ref{figure5}(b), underscoring the significance of two-particle physics in comprehending the many-body dynamics of the paired state. Furthermore, the similarity between the two curves is such that, by scaling the curve in Fig. \ref{figure5}(b) by a suitable factor near equal to the number of pairs $M/2$, one could achieve a close approximation of the curve presented in Fig. \ref{figure7}(g). This characteristic appears to be more than coincidental, considering the relation, demonstrated in the framework of the Richardson's integrability in Eq. 4 of Ref. [\onlinecite{combescot2011}], between the ground state energy of the many-body problem and the corresponding quantity in the single-pair problem.\\
The $X_2$ \textit{vs} $g$ curve depicted in Fig. \ref{figure7}(g) shows a quasi-linear dependence on the interaction strength $g$ when $g\lesssim 3$. The curve reaches a maximum around $g \approx 4$ and thereafter gradually decays for $g\gtrsim 5$. The Fourier analysis shown in Fig. \ref{figure7}(h) reveals a strong correlation between the behavior of the $X_2$ versus $g$ curve and the dependence on $g$ of the coefficients $A_1$ and $A_2$. These coefficients are respectively associated with a dominant periodicity of $\Phi_0$ or $\Phi_0/2$ in the Fourier expansion of the ground state, as described in Equation (\ref{eq:fourierSize}). A joint analysis of panels (g) and (h) of Fig.\ref{figure7} demonstrates that the peak of $X_2$, located at $g \approx 4$, corresponds to a maximum in $A_2$ and a rather complete suppression of $A_1$, so that the $\Phi_0/2$ periodicity is observed for $g>4$. For small interaction values ($g<3$), the quasi-linear behavior of the $X_2$ \textit{vs} $g$ curve is determined by the interplay between the dependence on $g$ of $A_1$ and $A_2$, while a $\Phi_0$ periodicity is observed. For these reasons, we conclude that a peak in $X_2$ \textit{vs} $g$ is observed in close vicinity of the crossover between $\Phi_0$ and $\Phi_0/2$ periodicity, which also marks the crossover between the behavior of a Cooper pairs superfluid and a gas of point-like composite bosons. The composite bosons phase, observed for $g\gtrsim 3$, is signaled by the suppression of the parity effect (see Fig. \ref{figure7}, panels (c) and (f)). On the other hand, parity effect is observed in the Cooper pairs superfluid phase, i.e. for $g\lesssim 2$.\\
Interestingly, well-developed Cooper-pairing superfluidity has been demonstrated in Ref. [\onlinecite{buccheri}] by studying the Josephson dynamics between weakly-coupled Richardson superconductors. According to Ref. [\onlinecite{buccheri}], a definite relative phase difference emerges even for very small numbers of pairs ($\sim 10$), so that BCS-like coherence phenomena are expected even in systems comprising a limited number of particles. Accordingly, the behavior of $X_2$, observed in Fig. \ref{figure7}(g), appears to be affected by a genuine BCS-BEC crossover, characterized by a peak in the $X_2$ \textit{vs} $g$ curve at the transition point.\\
In this context, a key clarification pertains to the observed $\Phi_0$ periodicity in the superfluid phase of Cooper pairs. The aforementioned periodicity does not indicate an absence of pairing. Rather, it arises when the system's size is comparable to the typical size of a Cooper pair. Consistent with expectations \cite{barash2008}, a period halving is achievable by increasing the system size while maintaining a constant interaction strength. In the bosonic phase, strong interactions lead to the formation of tightly bound pairs, with a typical size much smaller than that of the system. Consequently, a $\Phi_0/2$ periodicity becomes apparent in the bosonic phase. These observations suggest that size effects influence the BCS and BEC phases differently, thereby aiding in distinguishing these two regimes.

\section{Discussion and conclusions}
\label{sec:concl}
We have shown that the real-space BCS interaction, as introduced in Ref. [\onlinecite{rBCS}], can be understood as an infinite-range Penson-Kolb pairing mechanism, which coexists with an attractive Hubbard interaction. The proposed interaction enables the formation of fermion pairs through entirely non-local two-particle hopping processes, and the resulting Hamiltonian model retains its quantum mechanical nature for both weak and strong couplings. Analyzing the transport properties of this model is a challenging task, as it requires addressing non-equilibrium and strong correlation phenomena simultaneously. Rather than directly attempting to tackle this formidable problem, we have developed and discussed a Hamiltonian model that enables the extraction of various properties, including W. Kohn's charge stiffness, a parameter related to the system's conductivity.

The aforementioned model describes interacting fermions in a quantum ring under the influence of an external magnetic flux. We have shown that the flux enters the model in a highly intricate manner, simultaneously affecting both the kinetic and interaction terms. From an intuitive standpoint, much like the conventional Penson-Kolb model, even in the case of infinite range, the interaction part of the Hamiltonian can be understood as a pair hopping term. Consequently, the unconventional phase factor influencing the interaction term can be interpreted as the electromagnetic phase acquired by a Cooper pair during the infinite-range hopping process.

The resulting model exhibits Richardson integrability for both integer and half-integer values of the applied flux. For arbitrary flux values, we have employed DMRG and exact diagonalization methods to explore a genuine many-body problem influenced by a stationary current.

To emphasize the peculiarities and analogies of the proposed model with a more standard attractive Hubbard interaction, we conducted a thorough analysis of the two-particle problem. In this context, we have demonstrated that the real-space BCS interaction, unlike the attractive Hubbard interaction, produces a single bound state corresponding to a fermion pair. Excited states exist and correspond to unpaired fermions scattered by the interaction.

Analyzing the energy-flux curves for the ground state and the excited states, we observe $\Phi_0/2$ periodicity for the former and $\Phi_0$ for the latter, provided a sufficiently strong interaction strength is considered. A similar periodicity crossover is observed when varying the system size while keeping the interaction strength fixed. We have shown that this crossover, also present for the attractive Hubbard interaction, depends on the ratio between the pair's typical size $\xi$ and the system size $N$, with the periodicity $\Phi_0/2$ observed when $\xi/N \ll 1$.

The analysis of the charge stiffness for the real-space BCS interaction shows that the system's conductivity increases as the interaction strength increases for small or moderate values of the interaction strength $g$. For moderate or strong interaction values, the charge stiffness first reaches a maximum and then starts to decrease.

The above phenomenology, investigated in the case of the real-space BCS model, persists in the presence of many-body effects, which can be studied using DMRG simulations with a variable number of particles. In particular, we have demonstrated that the charge stiffness, and thus the system's conductivity, exhibits a similar behavior compared to that of the two-particle problem, confirming the special role of two-body physics in understanding the superconducting state. Moreover, the analysis of the charge stiffness as a function of the interaction strength, along with the study of the flux periodicity crossover, demonstrates that the observed behavior is related to a genuine BCS-BEC crossover. This crossover can be studied because size effects influence the BCS and BEC phases differently, thereby aiding in distinguishing these two regimes.

The analysis of the energy-flux curves does not reveal any evidence of a periodicity crossover towards a $\Phi_0/4$ periodicity, even for strong interaction values. Such a crossover would typically indicate the presence of Cooper's quartets. However, our observations suggest that the ground state of the system consists of Cooper pairs, aligning with a mean-field interpretation. Consequently, assuming that the theory adequately describes the system, we argue that the presence of Cooper quartets, as reported in Ref. [\onlinecite{quartets}], could be associated with a non-equilibrium condition where the external circuit provides the necessary energy to form a four-fermion aggregate. The formation of such a four-fermion aggregate would be suppressed in the ground state under equilibrium conditions, even in the presence of strong interactions.

Another option involves the possibility that Cooper quartets are stabilized within a non-trivial environment characterized by appropriate lattice structures, disorder, or quasi-periodicity. Interestingly, all of these mentioned conditions can be investigated by modifying the kinetic part of the model examined in this work. When subjected to these modifications, the model maintains Richardson's integrability for integer or half-integer values of the applied flux, thus providing an intriguing platform for studying exotic particle aggregates within the broader framework of BCS pairing.

Exploring these possibilities, and specifically, the conditions that promote the formation of Cooper quartets within the context of Richardson's integrability, remains a compelling and deserving subject for further attention and research.

\section*{Acknowledgment}
We gratefully acknowledge M. Salerno for enlightening discussions on interacting quantum systems. Our appreciation also goes to R. De Luca, R. Citro, M. Blasone, F. Illuminati and J. Settino for their insightful comments on various aspects of this study.

\appendix
\section{Simulating interacting fermions described by Eq.(\ref{eq:fewH}) by means of the Density Matrix Renormalization Group (DMRG) Method}
\label{app:A}
\begin{figure*}[h]
\includegraphics[scale=0.7]{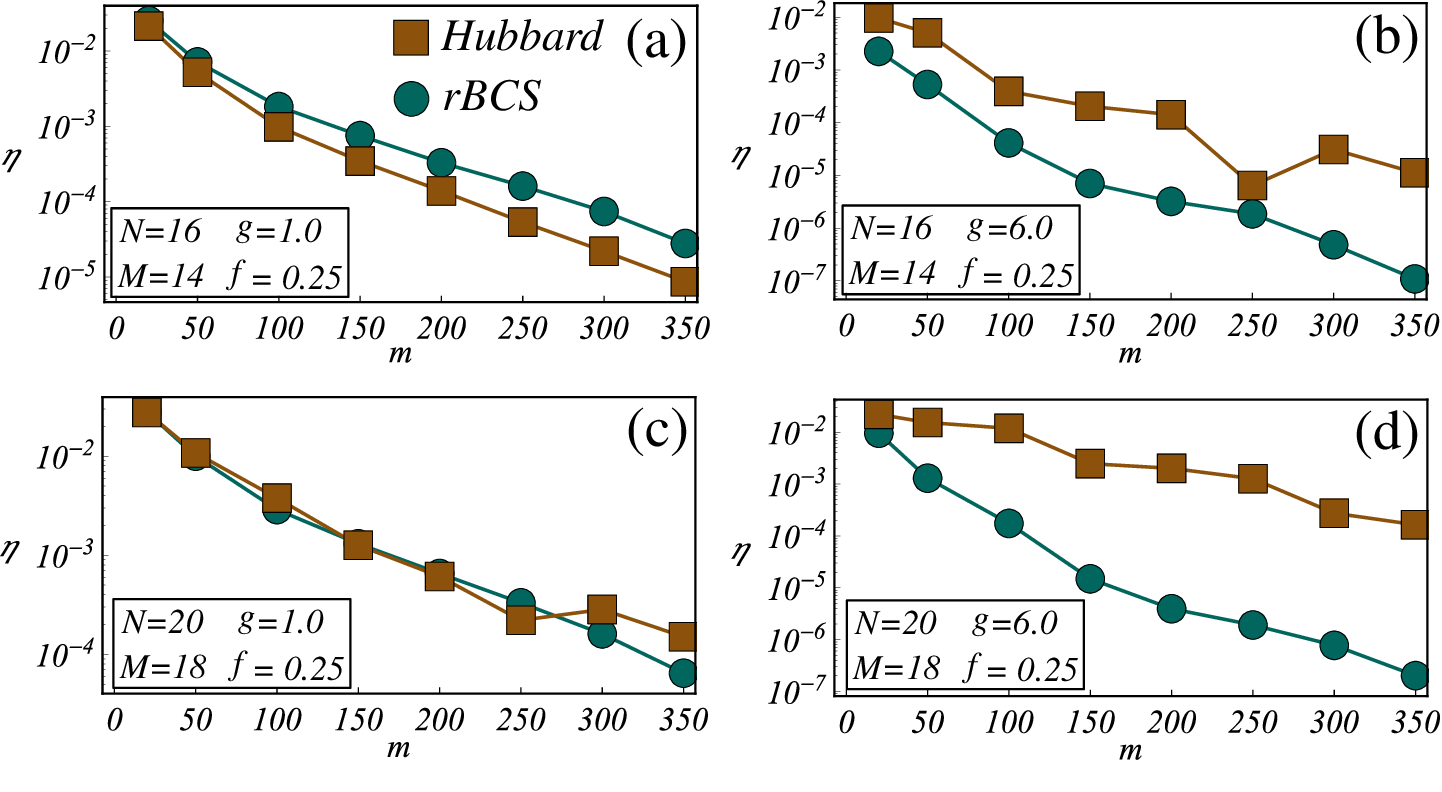}
\caption{Panels (a)-(d) illustrate the convergence of the ground state energy relative to the bond dimension $m$ in the matrix product state, as employed by the DMRG algorithm. To monitor the numerical accuracy, we define the convergence metric $\eta = |(E_{G}^{(m)} - E_{G}^{(400)}) / E_{G}^{(400)}|$, where $E_{G}^{(m)}$ represents the ground state energy calculated with bond dimension $m$. All simulations are performed with DMRG sweeps set to $5$. The legends within the figures detail the simulation parameters. The analysis reveals that for moderate interaction strengths (i.e., $g=1.0$, see panels (a) and (c)), the computational cost of simulating real space BCS interaction parallels that of the Hubbard interaction. However, in systems with strong interactions (i.e., $g=6.0$, see panels (b) and (d)), real space BCS interaction is simulated more efficiently than the attractive Hubbard model, a trend observed across various system sizes.}
\label{figure8}
\end{figure*}

To apply the DMRG algorithm, we adopt a spinless representation of the Hamiltonian as specified in Eq. (\ref{eq:fewH}). For this purpose, we map both the lattice positions and the spin degrees of freedom onto a unified lattice index via the operator transformation:
\begin{eqnarray}
\label{eq:map}
a_{\ell \uparrow} &\rightarrow& d_{2\ell-1}\nonumber\\
a_{\ell \downarrow} &\rightarrow& d_{2\ell},
\end{eqnarray}
where $\ell \in \{1,..., N \}$ denotes the original site index, while the newly introduced spinless operators follow standard anticommutation relations, i.e. $\{d_i,d_j^{\dag}\}=\delta_{ij}$ and $\{d_i^{\dag}, d_{j}^{\dag}\} =\{d_i, d_j\}=0$ with $i,j \in \{1,..., 2N \}$. Using the mapping in Eqs. (\ref{eq:map}), Eq. (\ref{eq:fewH}) is rewritten in the following form:
\begin{eqnarray}
\label{eq:spinlessH}
H=&-& t\sum_{\ell=1}^{N-1} \Big (d_{2\ell+1}^{\dag}d_{2\ell-1}+d_{2\ell+2}^{\dag}d_{2\ell} \Big)+h.c.\nonumber\\
&-& t e^{-i 2 \pi f} \Big (d_{1}^{\dag}d_{2N-1}+d_{2}^{\dag}d_{2N} \Big)+h.c.\nonumber\\
&-&\frac{g}{N}\sum_{\ell,r=1}^{N}d_{2\ell-1}^{\dag}d_{2\ell}^{\dag}d_{2r}d_{2r-1}.
\end{eqnarray}
The Hamiltonian presented in Eq. (\ref{eq:spinlessH}) characterizes the motion of spinless fermions within a lattice consisting of $2N$ lattice sites. The single-particle hopping is only allowed between lattice sites labelled by indices with the same parity (odd or even), being this selectivity reminiscent of the absence of spin-flipping mechanisms in the original Hamiltonian. Thus, the resulting terms describe a second-nearest neighbour hopping mechanism. Closed boundary conditions are enforced by flux-sensitive long-range hopping terms coupling lattice sites labelled by indices with the same parity. The pairing term couples fermions at sites of distinct parity, mirroring the BCS pairing mechanism where particles of opposite spin orientations are paired.\\
A relevant byproduct of this analysis is that, in virtue of the mapping in Eq. (\ref{eq:map}), the spinless Hamiltonian problem in Eq. (\ref{eq:spinlessH}) is also Richardson integrable for integer or half-integer values of the applied magnetic flux $f$, thus providing a further non-trivial example of integrable many-body quantum system.\\
The Hamiltonian in Eq. (\ref{eq:spinlessH}) is then converted into a spin operators language by implementing the Jordan-Wigner transformation:
\begin{eqnarray}
d_{j}=\prod_{k=1}^{j-1}\sigma_{z}^{(k)}\sigma_{-}^{(j)}\\
d^{\dag}_{j}=\prod_{k=1}^{j-1}\sigma_{z}^{(k)}\sigma_{+}^{(j)},
\end{eqnarray}
expressed in terms of Pauli's matrices $\sigma_{z}^{(k)}$ and $\sigma_{\pm}^{(k)}=\sigma_{x}^{(k)}\pm i \sigma_{y}^{(k)}$ acting on the lattice site labelled by $k$. The quantum states, empty or occupied, of the lattice site labelled by $k$ are specified by $|0\rangle_k$ or $|1\rangle_k$, respectively. The operators $\sigma_{\pm}^{(k)}$ act as rising/lowering operators on these states, so that $\sigma_{+}^{(k)}|0\rangle_k=|1\rangle_k$, $\sigma_{-}^{(k)}|1\rangle_k=|0\rangle_k$, $\sigma_{+}^{(k)}|1\rangle_k=\sigma_{-}^{(k)}|0\rangle_k=0$.\\
Using the Jordan-Wigner transformation, we get the spin Hamiltonian:
\begin{widetext}
\begin{eqnarray}
\label{eq:spinHam}
H &=& t \sum_{\ell=1}^{N-1}\Big(\sigma_{+}^{(2\ell+1)} \sigma_{z}^{(2\ell)} \sigma_{-}^{(2\ell-1)} + \sigma_{+}^{(2\ell+2)} \sigma_{z}^{(2\ell+1)} \sigma_{-}^{(2\ell)} \Big) + h.c.\nonumber\\
 &+& t e^{-i 2 \pi f} \Big [ \sigma_{+}^{(1)} \Big (\prod_{k=2}^{2N-2}\sigma_{z}^{(k)} \Big) \sigma_{-}^{(2N-1)} + \sigma_{+}^{(2)} \Big (\prod_{k=3}^{2N-1}\sigma_{z}^{(k)} \Big) \sigma_{-}^{(2N)} \Big ] +h.c.\nonumber\\
&-&\frac{g}{N} \sum_{\ell, r=1}^{N} \sigma_{+}^{(2\ell)} \sigma_{+}^{(2\ell-1)} \sigma_{-}^{(2r)} \sigma_{-}^{(2r-1)}.
\end{eqnarray}
\end{widetext}
The pairing term in Eq. (\ref{eq:spinHam}) can be compared with an attractive Hubbard interaction term, i.e. $H_U=-g \sum_{\ell}n_{\ell\uparrow}n_{\ell\downarrow}$, written in spin operators language:
\begin{eqnarray}
H_U &=&-g \sum_{\ell=1}^{N} \sigma_{+}^{(2\ell)} \sigma_{+}^{(2\ell-1)} \sigma_{-}^{(2\ell)} \sigma_{-}^{(2\ell-1)}=\nonumber\\
&=& -\frac{g}{4}\sum_{\ell=1}^{N} \sigma_{z}^{(2\ell-1)}\sigma_{z}^{(2\ell)}+\frac{g (N-2M)}{4},
\end{eqnarray}
being the comparison illustrative of the completely nonlocal nature of the real space BCS interaction in comparison with the local action of the attractive Hubbard term.\\
Since the original Hamiltonian in Eq. (\ref{eq:spinlessH}) preserves the total particles number $M=\sum_{j=1}^{2N}d_{j}^{\dag}d_{j}$, the same constraint, written in spin operators language, implies the conservation law:
\begin{eqnarray}
\label{eq:conservLaw}
\sum_{j=1}^{2N}\sigma_{z}^{(j)}=2(M-N).
\end{eqnarray}
The ground state of the Hamiltonian in Eq. (\ref{eq:spinHam}), complemented by the conservation law in Eq. (\ref{eq:conservLaw}), is an appropriate linear combination of the computational basis $|\{s\}\rangle=\bigotimes_{k=1}^{2N}|s_k\rangle_k$, with $s_{k} \in \{0,1\}$, and can be efficiently determined by the optimization of an open boundary matrix product state (MPS) through the DMRG algorithm, which is part of the ITensor library\cite{itensor}. Using this procedure, the many-body ground state
\begin{eqnarray}
|\Psi\rangle= \sum_{\{s\}}T^{s_1,...,s_{2N}}|s_1,..., s_{2N}\rangle
\end{eqnarray}
can be approximated by using the following tensor train decomposition\cite{rizzi}:
\begin{eqnarray}
T^{s_1,...,s_{2N}}=\sum_{\{ \alpha \}}A_{\alpha_1}^{s_1}A^{s_2}_{\alpha_1,\alpha_2} ... A^{s_{2N-1}}_{\alpha_{2N-1},\alpha_{2N}}A_{\alpha_{2N}}^{s_{2N}},
\end{eqnarray}
written in terms of three-index tensors $A^{s_k}_{\alpha,\beta}$, having a physical index $s_k$ and a pair of bond indices $\alpha, \beta \in \{1,..., m\}$. The bond dimension $m$ governs the expressivity of the MPS representation, which becomes increasingly accurate as $m$ grows. Typically, representing a tensor with $N$ indices, each of dimension $d$, demands $d^N$ parameters. In contrast, an MPS approximation with bond dimension $m$ reduces this to only $Ndm^2$ parameters, limiting the parameter growth from exponential to linear with respect to $N$. The MPS framework is particularly successful in capturing the ground state of one-dimensional gapped systems with local interactions, a feature associated with the exponential decay of correlation functions or, equivalently, to the area law of entanglement entropy\cite{arealaw}. Consequently, even with a limited bond dimension $m$, an MPS can closely approximate the exact ground state of such systems.\\
Considering the all-to-all interaction detailed in Eq. (\ref{eq:spinHam}), one might wonder if the ground state of a system governed by such a Hamiltonian can be efficiently approximated with an MPS. Despite the complete non-locality induced by the pairing term, it is possible to demonstrate that the numerical algorithm retains its effectiveness.\\
This statement is supported by detailed numerical experiments, with selected results shown in Fig. \ref{figure8}. Using the DMRG algorithm, we have studied the convergence of the ground state energy as a function of the bond dimension $m$, while keeping the number of sweeps at $5$ and the applied flux at $f=0.25$. The convergence, for both the Hubbard and the real-space BCS interactions, has been monitored by introducing the metric $\eta=|(E_{G}^{(m)} - E_{G}^{(400)}) / E_{G}^{(400)}|$, measuring the relative deviation of the ground state energy $E_{G}^{(m)}$ computed with bond dimension $m$ with respect to the reference value $E_{G}^{(400)}$. We have examined two system sizes, namely $N=16$ (with $M=14$) and $N=20$ (with $M=18$), and two interaction strengths, i.e. $g=1.0$ and $g=6.0$. Results obtained by setting the same system size belong to the same raw of the figure matrix reported in Fig. \ref{figure8}, while those obtained with the same interaction strength belong to the same column. Exponential decay of the $\eta$ versus $m$ curves is invariably observed, indicating good convergence properties of the algorithm. The analysis also reveals that for moderate interaction strength (i.e., $g=1.0$, see panels (a) and (c)), the computational cost of simulating real space BCS interaction parallels that of the Hubbard interaction. Surprisingly, in systems with strong interactions (i.e., $g=6.0$, see panels (b) and (d)), real space BCS interaction is simulated more efficiently than the attractive Hubbard model, a trend observed across various system sizes. This result can be understood as the effect of approaching the infinite-interaction limit of the real space BCS model, an integrable model\cite{kerman,ormand,pehlivan} with a particularly simple structure of the ground state. In the latter circumstance, due to the all-to-all nature of the interaction, each system site becomes equivalent to any other, thus providing a drastic reduction of the information to be encoded by the MPS. Undoubtedly, the uniform character of the all-to-all interaction is here determinant in achieving the observed results. These findings well align with recent studies\cite{searle24, bettaque}, which have demonstrated that uniform all-to-all spin interactions can be efficiently represented using matrix product operators and simulated with tensor network strategies. Furthermore, the extensive body of research\cite{braun} on Richardson integrable systems demonstrates that DMRG simulations frequently produce results that closely approximate exact solutions.\\
In conclusion, real-space BCS interaction proves to be particularly suited for DMRG simulations, offering a computational cost that is similar to, if not better than, that of the attractive Hubbard interaction.

\section{Scaling Behavior of Charge Stiffness with the System Size}
\label{app:B}
\begin{figure*}[h]
\includegraphics[scale=0.7]{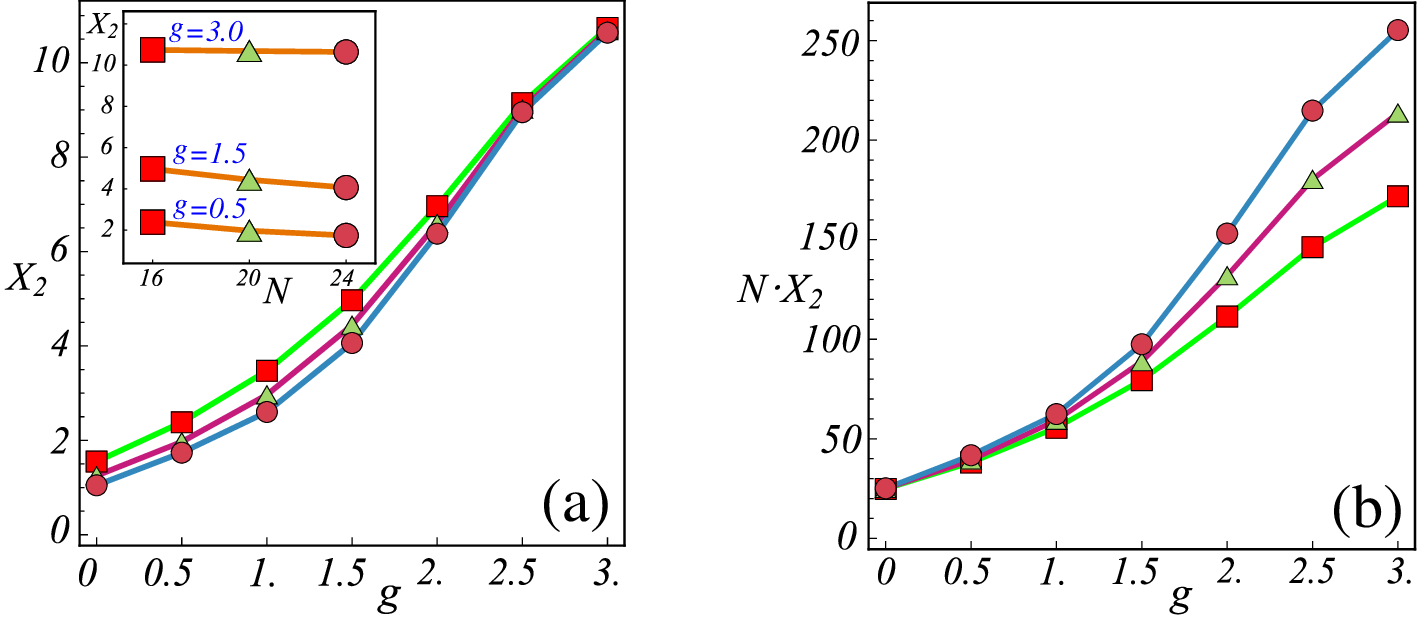}
\caption{(a): $X_2$ versus $g$ curves obtained for systems of different sizes $N$ close to half-filling ($M/(2N)\approx 0.45$). Specifically, square symbols ($\Box$) represent a system size of $N=16$ with a particle number of $M=14$; triangle symbols ($\triangle$) indicate $N=20$ and $M=18$; circle symbols ($\bigcirc $) represent $N=24$ and $M=22$. The inset displays the $X_2$ versus $N$ curves for interaction values noted within the figure. (b): $N \cdot X_2$ versus $g$ curves obtained for the same $N$ and $M$ values used in panel (a). For small to moderate interaction strengths (i.e., $g<1$), the curves are observed to collapse onto each other. The real-space BCS interaction has been considered for both the panels.}
\label{figure9ab}
\end{figure*}

In the following analysis, we examine the real-space BCS interaction and investigate how the charge stiffness $D_c$ (Eq. (\ref{eq:stiff}) in the main text) scales with the system size $N$. Since charge stiffness is proportional to $X_2$ (Eq. (\ref{eq:x2}) in the main text), examining $X_2$ sheds light on $D_c$. This analysis is presented in Fig. \ref{figure9ab}(a), which shows the variation of $X_2$ with the interaction strength $g$ for different system sizes $N \in \{16, 20, 24\}$. Here, square ($\Box$), triangle ($\triangle$), and circle ($\bigcirc$) symbols denote system sizes of $N=16$ with $M=14$, $N=20$ with $M=18$, and $N=24$ with $M=22$, respectively. We focus on fermionic systems near half-filling ($M/(2N)\approx 0.45$) and explore interaction strengths $g<3$ to characterize the parameter region containing the BCS regime. The $X_2$ versus $g$ curves, which are growing functions of $g$, suggest enhanced conduction properties under stronger interactions, echoing observations made in Fig. \ref{figure7}(g) of the main text. An inset in Fig. \ref{figure9ab}(a) presents $X_2$ versus $N$ curves at fixed $g$ values, indicating that $X_2$ is a decreasing function of $N$, as expected. In virtue of the proportionality relation $D_c \propto N X_2$, a correct picture of the charge stiffness $D_c$ is obtained by studying the quantity $N \cdot X_2$. Fig. \ref{figure9ab}(b) displays $N \cdot X_2$ versus $g$ curves for various $N$, showing that curves for $g<1$ overlap, whereas those obtained for $1<g<3$ exhibit larger $D_c$ values in bigger systems, hinting that the thermodynamic limit of $N \cdot X_2 \propto D_c$ is not yet achieved.


\begin{thebibliography}{99}
\bibitem{nota1} Notably, the real-space BCS Hamiltonian, denoted as $H_{rBCS}$, can also be interpreted as the cumulative interaction energy arising from the presence of an attractive Hubbard term ($H_U$), a Penson-Kolb term ($H_{PK}$), and a residual long-range pair-hopping term ($H_{LR}$), thus yielding $H_{rBCS}=H_{U}+H_{PK}+H_{LR}$.
\end{thebibliography}
\end{document}